\documentclass[twocolumn,letter,prl,showpacs]{revtex4}
\usepackage[latin9]{inputenc}
\usepackage{amsmath}
\usepackage{amssymb}
\usepackage{graphicx}
\usepackage{esint}

\makeatletter

\usepackage{color}\usepackage{dcolumn}\usepackage{bm}\usepackage{epsfig}

\makeatother

\begin{document}

\title{Photonic Bound State in the Continuum for Strong Light-matter Interaction}

\author{Chang-Ling Zou}

\author{Jin-Ming Cui}

\author{Fang-Wen Sun}
\email{fwsun@ustc.edu.cn}

\author{Xiao Xiong}

\author{Xu-Bo Zou}

\author{Zheng-Fu Han}

\author{Guang-Can Guo}

\affiliation{Key Lab of Quantum Information, University of Science and Technology
of China, Hefei 230026}
\begin{abstract}
The photonic bound state in the continuum (BIC) is discovered in a hybrid photonic circuit with low refractive index waveguide
on a high refractive index thin membrane, where the optical dissipation is forbidden because of the destructive interference
of different leakage channels. Based on the photonic BIC, the low mode area in a hybrid waveguide and high quality factor in a
microresonator can be applied to enhance the light-matter interaction. Taking the fabrication-friendly polymer structure on diamond membrane as
an example, those excellent optical performances can exist in a wide
range of structure parameters with large fabrication tolerance and
induce the strong coupling between photon and nitrogen-vacancy center
in the diamond for scalable quantum information processors and networks.
Such a fabrication-friendly structure with photonic BIC is also very promising in laser, nonlinear optical and
quantum optical applications.
\end{abstract}

\pacs{42.55.Sa, 05.45.Mt, 42.25.-p,42.60.Da}

\maketitle
\emph{Introduction.- }Great progresses have been achieved recently
in the photonic integrated circuits (PIC) for classic and quantum
information processes \cite{saleh2007fundamentals,o2009photonic}.
Efforts have been dedicated to exploit new materials, such as materials
with excellent nonlinear optical properties \cite{peyronel2012quantum,eggleton2011chalcogenide,kim2012nonlinear,jacobsen2006strained}
and solid quantum emitters \cite{balasubramanian2009ultralong,koehl2011room},
for better optical performances and functionality. And a range of
new physical phenomena has been studied in PIC and applied to develop
new components, including the photonic Aharonov-Bohm effect \cite{fang2012photonic}
and topological insulator \cite{khanikaev2013photonic,rechtsman2013photonic}.
In PIC, photons are processed by the basic components composed of
waveguide and resonator structures, where high refractive index (RI)
contrast to environment is essential. It has been long known that
the light in low RI contrast structure will leak through coupling
to continuum modes in substrate. This can be intuitively explained
by quantum physics that the wave cannot be localized upon a potential
well. Thus, the materials for PICs are restricted to
high RI. However, fabrication difficulties are often encountered for some high
RI materials, such as the diamond whose optical performances suffer
from the surface scattering caused by sidewall roughness \cite{faraon2011resonant,hausmann2012integrated}.

It is instructive to consider the low RI contrast PIC. Actually, the
bound states do exist upon potential well. This intriguing phenomenon,
the so-called bound state in the continuum (BIC), was first proved
by Von Neumann and Wigner in 1929 \cite{Neumann}, and has been extensively
studied recently in electronic and photonic structures \cite{capasso1992observation,marinica2008bound,diaz2009extraordinary,molina2012surface,plotnik2011experimental}.
In this Letter, we propose a novel photonic waveguide based on the
BIC, which is made by low RI material on a high RI substrate, without
specially engineered potential distribution \cite{molina2012surface}
or multiple waveguides by the virtue of symmetry \cite{plotnik2011experimental}.
In such a hybrid waveguide, optical dissipating is forbidden because
of the destructive interference of different dissipation channels
under certain geometric parameters. Remarkably, the BIC is also demonstrated
in the rotating-symmetry microring structure, allowing ultrahigh quality
factor resonances. Therefore, our study brings the BIC from the basic
scientific interest of fundamental quantum physics to practical applications,
permiting a new kind of PIC. In this work, we
take the diamond as an example, for the reason that photonic structure
is very demanded to enhance the emission and collection of zero-phonon
line (ZPL) of nitrogen-vacancy center (NVC) in diamond, then to improve
the fidelity in quantum control \cite{ladd2010quantum,maurer2012room}
and resolution in magnetic and biologic sensing\cite{mamin2013nanoscale,kucsko2013nanometer}
with NVC. By combining with the NVCs and the fabrication-friendly polymer
photonic structures on a diamond membrane, strong light-matter interaction
is promising. The BIC based diamond chip provides an excellent platform
for integrated quantum networks \cite{aharonovich2011}.

\begin{figure}
\includegraphics[width=7.5cm]{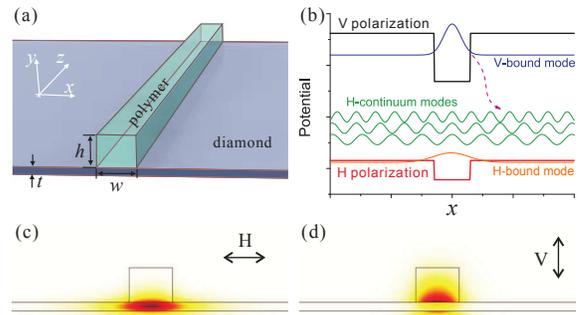} \caption{(color online) (a) The hybrid photonic BIC structure with polymer
waveguide on a thin diamond membrane. (b) The effective potential
for modes. (c) and (d) Field distributions of Horizontal or Vertical
polarized modes bounded to the waveguide, with polymer waveguide height
$h=400\mathrm{nm}$, width $w=500\mathrm{nm}$, and membrane thickness
$t=100\mathrm{nm}$. }

\label{f1}
\end{figure}

\emph{Bound State in the Continuum.-} The light propagating in waveguide
along $z$-axis with slow-varying approximation can be described by
the Schrödinger equation
\begin{equation}
i2\beta\partial_{z}\psi(\mathbf{r})=[\nabla^{2}-U(\mathbf{r})]\psi(\mathbf{r}),
\end{equation}
by replacing the time $t$ by $z$. Here, $\psi(\mathbf{r})e^{i\beta z-i\omega t}$
is the scalar wave function, $\omega$ and $\beta$ are respectively
the light frequency and propagation constant. The potential $U(\mathbf{r})=\beta^{2}-n^{2}(\mathbf{r})k^{2}$,
with $n(\mathbf{r})$ is the dielectric RI and $k=2\pi/\lambda$ is
the vacuum wavenumber. Usually, waveguide structures are composed
of high RI waveguides in low RI environment, where the guiding modes
satisfy $U(\mathbf{r})>0$ in environment and the modes are well confined
with only evanescent field outside. However, in the case of low RI
waveguide in high RI environment, the mode energy is higher than the
potential of environment as $U(\mathbf{r})<0$. The guiding mode coupling
to continuum modes will lead to significant leaking loss.

To construct PICs beyond the limitation of high RI contrast, we consider
the hybrid structure consisting of polymer waveguide on an air-suspended
diamond membrane as depicted in Fig. 1(a). In this film-like substrate,
wavefunctions are confined along $y$-direction, and the engineered
continuum is reduced to be one dimensional (along $x$). Under the
two-dimensional approximation by treating the regions with and without
polymer both as film waveguides separately, the Schrödinger equation
reduced to one-dimension as \citep{zaghlous1999simple}
\begin{equation}
i\beta\partial_{z}\psi(x,z)=[\partial_{x}^{2}-U(x)]\psi(x,z).
\end{equation}
Here, $n(x)$ is replaced by the effective 2D RI. In the following,
we focus on the working wavelength $\lambda=637\mathrm{nm}$ (ZPL
of NVCs), with RIs of the polymer and diamond are $n_{p}=1.49$ and
$n_{d}=2.41$, respectively. The effective RIs of the region with and
without polymer for the Horizontal (Vertical) polarization are $n_{1,H(V)}=1.879(1.536)$
and $n_{2,H(V)}=1.820(1.171)$, which are obtained as the effective
mode indices $(n_{eff}=\beta/k)$ of fundamental propagation modes of
the film waveguide. As shown in Fig.$\ $\ref{f1}(b), the external
polymer waveguide induces a potential well which gives rise to bound
modes. Figures$\ $\ref{f1}(c) and \ref{f1}(d) are the numerically
solved fundamental bound modes with different polarization. Since the
depth of the potential well for horizontal-polarized bound mode (HBM)
is much lower than that for vertical-polarized bound mode (VBM), the
HBM is weakly confined in the membrane and widely distributed along
$x$-direction, while the VBM is strongly confined laterally. This
result suggests that very large loss of HBM for bending waveguides
in practical applications. Therefore, we restrict the study to VBM
in the following.

\begin{figure}
\includegraphics[width=7.5cm]{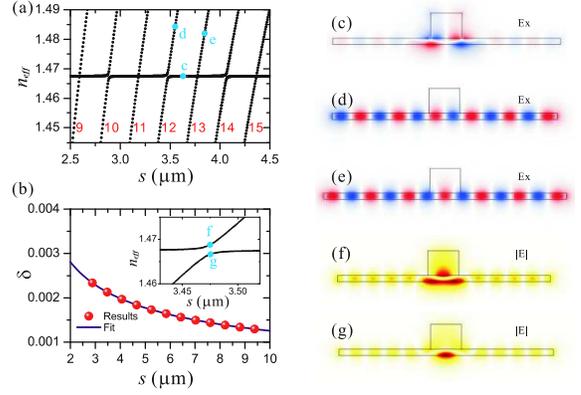} \caption{(color online) (a) The effective mode indices of VBM and HSMs with
varying slab width $s$. (b) The minimal effective mode index differences
at anticrossing. Inset: the anticrossing for $m=12$. (c)-(g) Field
distributions of $s$ denoted in (a) and (b), with same $h$, $w$ and
$t$ in Fig.\ref{f1}.}

\label{f2}
\end{figure}

It is intuitively obvious that the VBM can dissipate through coupling
to infinite horizontal-polarized continuum modes (HCM) above the potential
barrier, due to the energy of VBM is much higher than the potential
barrier of HBM {[}Fig.$\ $\ref{f1}(b){]}. The Hamiltonian for the
VBM in continuum modes can be written as ($\hbar c=1$)
\begin{equation}
H=\beta_{b}b^{\dagger}b+\sum_{m}\beta_{m}c_{m}^{\dagger}c_{m}+\sum_{m}(g_{m}c_{m}^{\dagger}b+h.c.),
\end{equation}
with $b$ and $c_{m}$ are respectively annihilation operators of
the VBM and $m$-th HCM, $\beta_{b}$ and $\beta_{m}$ are the corresponding
propagation constants, $g_{m}$ is the coupling strength.

\emph{BIC in Hybrid Waveguide.-} We firstly study the coupling between
VBM and HCMs of a slab with finite width $s$. In this case, HCMs
are discrete under the quantization condition $k_{x}s\approx2\pi m$,
with the wave-vector along $x$-direction $k_{x}=\sqrt{n_{2,H}^{2}-n_{m}^{2}}kw$.
The integer $m$ is the order of HCM and $n_{m}=\beta_{m}/k$ is the
effective mode index. Figure$\ $\ref{f2}(a) shows the numerically
solved effective mode indices of eigenmodes in the hybrid waveguide,
where $n_{m}\approx\sqrt{n_{2,H}^{2}-(2\pi m/ks)^{2}}$ changes with
$s$ while the effective mode index of VBM $n_{b}=\beta_{b}/k$ is
constant. When VBM and HSMs are near resonance $(n_{m}\approx n_{b}$),
there are two different behaviors in the trajectories of the mode
index against $s$: crossing when $m$ is odd and anti-crossing when
$m$ is even. To interpret these results, the profiles of horizontal
component of electric field ($E_{x}$) for VBM and HSMs are plotted
in Fig.$\ $\ref{f2}(c)-\ref{f2}(e). Although the electric field
of VBM is quasi-V-polarized, its $E_{x}$ is nonzero at corners due
to diffraction {[}Fig.$\ $\ref{f2}(c){]}. The parity
of $E_{x}$ of VBM is odd while parity of $m$-th HSM is determined
by $m$ {[}Figs.$\ $\ref{f2}(d) and \ref{f2}(e){]}. When $m$ is
odd, the VBM is orthogonal with $m$-th HSM and their mode index trajectories
cross each other without interplay. However, for even $m$, the
overlap between wave-functions of VBM ($\varphi_{b}$) and $m$-th
HSM ($\varphi_{m}$) will induce coupling. Due to the interaction,
the eigenmodes of this hybrid waveguide are hybridization of VBM and
HSMs, showing an effective index difference between the new eigenmodes
are $\delta=\sqrt{4g_{m}^{2}+\triangle^{2}}/k$, with $\triangle=(n_{m}-n_{b})k$
is the detuning. It means that nonzero $g_{m}$ leads to the anti-crossing
that $\delta$ is always greater than $0$. As an example, a detailed
illustration of anti-crossing between VBM and $12$-th HSM is shown
in the inset of Fig.$\ $\ref{f2}(b). When the two modes are on-resonance
($n_{12}=n_{b}$), the splitting between two eigenmode branches reaches
the minimum ($\delta=2g_{12}/k$), with hybrid eigenmodes shown in
Figs.$\ $\ref{f2}(f) and \ref{f2}(g).

\begin{figure}
\includegraphics[width=7.5cm]{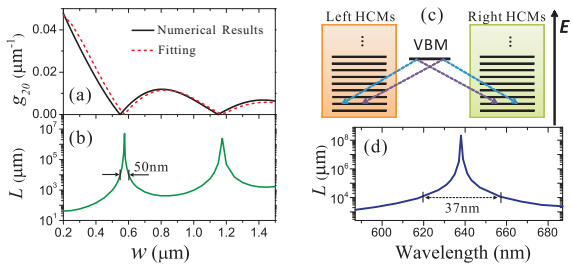} \caption{(color online) (a) The coupling strength $g_{20}$ against waveguide width $w$. The
solid line is the numerical simulation result, and the dashed line
is the analytical result. (b) The propagation length of the bound
waveguide modes for $s=\infty$ against the waveguide width with $\lambda=637\mathrm{nm}$.
(c) The schematic illustration of the BIC for
the VBM coupled to the HCMs. (d) The calculated propagation length
$L$ of VBM for $s=\infty$ against wavelength, with $w=575\mathrm{nm}$.}

\label{f3}
\end{figure}

The coupling strength $g_{m}$ depends on the normalized field
overlap between VBM and HSM which is approximately proportional to
$1/w\sqrt{s}$. As shown in Fig. \ref{f2}(b), the minimums $\delta=2g_{m}/k$
which are deduced from Figs.$\ $\ref{f2}(a) are plotted and consistent
with the fitting by $1/\sqrt{s}$. There are two channels
for VBM dissipating to each continuum mode as diffracting at corners
of waveguide. As a consequence, we have
\begin{equation}
g_{m}\approx\widetilde{g}_{m}\times\frac{|1+e^{ik_{x}w}|}{2w\sqrt{s}},\label{e5}
\end{equation}
which contains the interference of two channels with paths phase difference
$k_{x}w\approx\sqrt{n_{2,H}^{2}-n_{m}^{2}}kw$, and $\widetilde{g}_{m}$
is constant. This means that the dissipation of VBM can be totally
frozen ($g_{m}=0$) by carefully choosing geometry of polymer waveguide:
when $\mathrm{cos(}k_{x}w)=-1$ is achieved, the dissipation cancels because of the destructive interference, leading to a photonic
BIC. To confirm this conclusion, we calculate $g_{20}$ as a function
of $w$ numerically from anti-crossings $g_{m}=\delta k/2$. As displayed
in Fig.$\ $\ref{f3}(a), the data are consistent with Eq. (\ref{e5}),
showing oscillations and approaching zero at certain $w$.

Since there are infinite continuum modes when $s$ goes infinite,
the membrane serves as a reservoir and the dynamics of VBM is Markovian.
Thus, the energy of VBM exponentially decays with propagation distance
as $I(z)=I(0)e^{-z/L}$ with the decay length $L=\sqrt{n_{2,H}^{2}-n_{b}^{2}}/2g_{b}^{2}sn_{b}$
\cite{zoudd}. Here, $g_{b}$ is the coupling strength between VBM
and near resonance HSM ($n_{m}\approx n_{b}$), and $g_{b}^{2}s\approx[\widetilde{g}_{m}\mathrm{cos(}\frac{k_{x}w}{2})/w]^{2}$
is independent on $s$. Figures \ref{f3}(b) and \ref{f3}(d) show
the dependence of $L$ on $w$ and wavelength. These results indicate
that $L$ can be larger than $1\mathrm{m}$, corresponding to the
BIC that VBM is almost totally decoupled from the continuum modes.

\begin{figure}
\includegraphics[width=7.5cm]{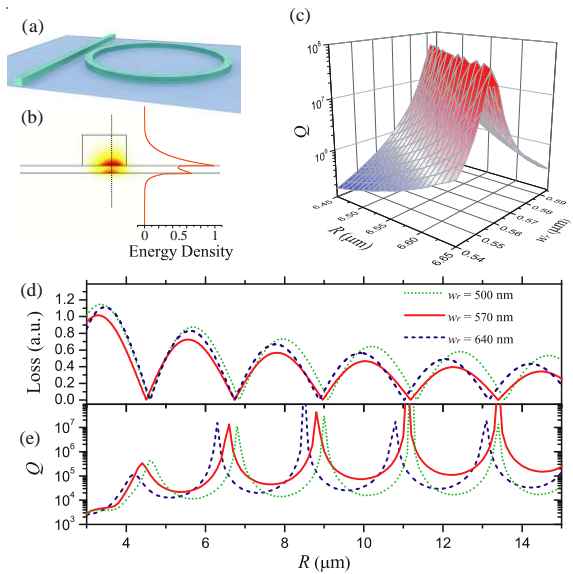} \caption{(color online) (a) The schematic of the hybrid microring coupling
to a hybrid waveguide. (b) The energy density distribution at the
cross section of the hybrid microring. The solid curve is the field
distribution along $y$-direction, as indicated by the dashed line.
(c) The tolerance of the \emph{Q} factor versus the $R$ and microring
width $w_{r}$. (d) and (e) The predicted loss and calculated \emph{Q}
factor of the microring against the radius $R$, with different microring
width.}

\label{f4}
\end{figure}

\emph{BIC in Hybrid Microresonator.- }Microresonator is another important
component in PIC, which can trap light and
enhance the light-matter interaction. Figure$\ $\ref{f4}(a) is a
typical microring resonator coupled to a waveguide, whose geometry is
characterized by the width of cross section ($w_{r}$) and exterior
radius ($R$). Whispering gallery modes (WGMs) in such hybrid microring
are solved numerically by an axis-symmetry model, with a perfect match
layer applied to calculate the quality factor ($Q$) accurately. For
the vertical-polarized WGMs, similar to the waveguide VBM, the bound
modes will couple to the continuum modes inevitably, which will induce
extra energy loss and limit the $Q$ of WGMs. In such axis-symmetry
structure, the WGMs couple to the horizontal-polarized cylindrical
waves in membrane $J_{q}(nkr)e^{iq\phi}$, where $J_{q}(z)$ is Bessel
function of order $q$ \cite{ZouOE13}. However, since the mode profile
of WGMs tends to outer corner {[}Fig.$\ $\ref{f4}(b){]}, the two
passages of dissipation (diffraction at two corners) are unbalanced.
The loss of bound modes to continuum is proportional to $|J_{q}(nkR)-\xi J_{q}(nk(R-w_{r}))|$,
where the factor $\xi<1$ accounts for the asymmetric coupling strengths
at two corners. In Figs.$\ $\ref{f4}(d) and \ref{f4}(e), the leaking
loss and $Q$ of WGMs with wavelength around $637\mathrm{nm}$ is
calculated against $R$. The prediction is consistent with numerical results. The oscillation of dissipation loss and $Q$
is due to the standing-wave behavior of cylindrical Bessel wave in
the radius direction. There are several peaks of $Q$ greater than
$10^{6}$ in Fig.$\ $\ref{f4}(e), indicating the existence of BIC, i.e.
the two dissipation channels cancel each other for specific $R$s.
It is worth noting that the horizontal-polarized WGMs are not found
in our numerical simulations, due to their extremely low $Q$ caused
by very large bending loss of HBM.

\emph{Strong Coupling.-} Utilizing the Purcell effect \cite{Purcell},
the coherent ZPL emission of NVC embedded in diamond will be enhanced
by interacting with bound photonic modes. For example, the effective
mode area of hybrid waveguide is $A=0.092\mathrm{\mu m^{2}}$ for
$w=570\mathrm{nm}$, corresponding to enhanced spontaneous emission
with Purcell factor $F=2.74$ and waveguide collection efficiency
of ZPL $\zeta=F/(F+1)=0.73$. In addition to the enhanced emission
and collection, the undesired phonon sidebands emission of NVC can
be suppressed as the BIC can only exist around ZPL {[}Fig.$\ $\ref{f3}(d){]}.

For the microring with $w_{r}=0.57\mathrm{\mu\mathrm{m}}$ and $R=6.6\mathrm{\mu\mathrm{m}}$,
the $Q$ of WGM is as high as $6.5\times10^{7}$ for BIC. The mode
volume for NVC in diamond $V=1.89\mathrm{\mu m^{3}}$ can be approximated
by $\pi RA$ as the mode profile of WGMs at the cross section is close
to that of waveguide. In such a high $Q/V$ hybrid microresonator,
the Purcell factor of ZPL can be as large as $F=\frac{3}{4\pi^{2}}\frac{Q}{V}(\frac{\lambda}{n_{d}})^{3}=828$.
The maximum coherence coupling strength between photon and zero-phonon
transition of NVC is $G=\sqrt{\frac{\omega d^{2}}{2\hbar\varepsilon V}\times\eta}\approx242\times2\pi$
MHz, where $d$ is the dipole momentum and $\eta=0.05$ is the percentage
of ZPL in the total emission of NVC. As the decay rate of NVC $\gamma\approx13\times2\pi$
MHz, the strong coupling ($G>\gamma,\kappa$) can be achieved when
the cavity dissipation rate $\kappa=\omega/2Q<G$, i.e. $Q>10^{6}$.
For practical experiment, the dependence of $Q$ on the variance of
$R$ and $w_{r}$ is studied in Fig.$\ $\ref{f4}(c), showing good
fabrication tolerance of BIC for strong coupling.

\emph{Discussion.-} Based on the BIC, the proposed structure is feasible
in experimental demonstration. The single crystal diamond membrane can be prepared by chemical methods \citep{fairchild2008fabrication,aharonovich2012homoepitaxial},
and its surface roughness can be reduced to a few nanometers by mechanical
polishing or phonon-assisted etching \citep{yatsui2012realization}.
As for the polymer nanostructures, which possess very excellent optical
properties and are compatible to other materials \citep{dong2010high},
they can be fabricated via mature nano-fabrication techniques, such
as electron beam lithography, nano-imprint and direct laser writing.
In addition, the BIC shows fabrication tolerance of wide range of
parameters. For example, ensuring $L>10$ $\mathrm{mm}$, the fabrication
tolerance of $w$ and the wavelength bandwidth are about $50$$\mathrm{nm}$
and $37\mathrm{nm}$, respectively. It worth noting that the diamond
membrane is not necessary air-suspended, it can be supported on a low RI
polymer or silica substrate.

Such a photonic BIC structure has shown nice optical performance for
diamond-based photonic application and various applications with NVC.
The general principles of BIC can also be extended to other low
RI material on high RI substrate. For
the photonic BIC waveguide with nonlinear optical materials,
the optical mode with low effective area can be use to enhance
the nonlinear interaction, which is very promising
in the chip-scale nonlinear optical application with novel
optical phenomena. Lasers can be generated with the
microring structure on the optical gain material with
simple design. The photonic BIC also shows potentials
in the application of new optical material without
complicated nanophotonic fabrication.

\emph{Conclusion.-} We have proposed and demonstrated the
photonic BIC in low RI waveguide and microring structures
on a high RI substrate. The BIC based PIC is feasible for the experiments.
The circuit with polymer on diamond membrane with low
loss BIC has fabrication tolerance of about $50\mathrm{nm}$. Strong coupling between ZPL transition of NVC and photon is possible,
offering an excellent platform for future scalable quantum information
processor and networks. Such a photonic BIC will also benefit many
chip-scale optical applications in classic and quantum optics.

\textbf{Acknowledgements} This work was supported by the National
Fundamental Research Program of China (No.2011CB921200), the Knowledge
Innovation Project of Chinese Academy of Sciences (No.60921091), National
Natural Science Foundation of China (No.11004184) and NCET.

\end{document}